\documentclass[preprint,showpacs,superscriptaddress,amsmath,amssymb]{revtex4}
\usepackage{dcolumn}
\begin{document}

\preprint{RPC}

\title{Relations Between Coefficients of Fractional Parentage}
\author{L. Zamick}
\affiliation{Department of Physics and Astronomy, Rutgers University, 
Piscataway, New Jersey 08854, USA }

\date{\today}

\begin{abstract}
For each of the (9/2), (11/2) and (13/2) single j shells we have only one 
state with J=j $v$=3 for a five particle system.  For four identical particles 
there can be more than one state of seniority four.  We note some ``ratio'' 
relations for the coefficients of fractional parentage for the four and five 
identical particle systems, which are found in the works of de Shalit and 
Talmi [1,2] to be useful for explaining the vanishing of a five particle cfp.  
These relations are used to show that there is a special (g$_{9/2}$)$^{4}$ 
I=4 $v$=4 wavefunction that cannot be admixed with an I=4 $v$=2 wavefunction, 
even with seniority violating interactions.
\end{abstract}

\maketitle

\section{Introduction}
In examining the Bayman-Lande tables [3] of coefficients of fractional 
parentage (cfp) we note some interesting relationships between (cfp)'s 
involving five nucleons with seniority three and four nucleons with seniority 
four.  These results are of particular interest for the $g_{9/2}$, $h_{9/2}$, 
and $i_{13/2}$.  For each of these shells there is only one state with total 
angular momentum J=j and seniority $v$=3.  For the four particle system for 
certain angular momenta I there are in some cases more than one state of 
seniority four.  For example for the $(g_{9/2}^{4})$ configuration there are 
two ($v$=4 I=4) states and two ($v$=4 I=6) states.  For $(i_{13/2})^{4}$ 
there are three ($v$=4 I=7) states, etc.

Let us designate the different $v$=4 states by $v$$_{i}$.  We define

\centerline{$C(v_{i}) = [j^{3}(J_{o}=j) j| j^{4} I v_{i}]$} 

\noindent and 

\centerline{$D(v_{i}) = [j^{4} Iv_{i}j| j^{5} (J=j)v=3]$}

\noindent where we use the standard notation for coefficients of fractional 
parentage.
 
The following relationships should be noted.

\begin{equation}
\frac{C(v_{i})}{C(v_{j})} = \frac{D(v_{i})}{D(v_{j})} 
\end{equation}

These relations have been explained in the books of deShalit and Talmi [1] and 
Talmi [2].  See especially Eq. (19.31) of ref. [2].

\begin{eqnarray} 
&&[j^{v+1}(v + 1,\alpha_{1}J_{1})jJ|\}j^{v + 2}v \alpha J] 
\nonumber\\
&=& (-1)^{J+j-J_{1}} \sqrt{\frac{2(2J_{1}+1)(v + 1)}{(2J + 1)(v + 2)(2j+
1-2v)}} \nonumber \\
&\times& [j^{v} (v \alpha J)jJ_{1}|\} j^{v + 1}(v + 1, \alpha_{1}) 
J_{1}] 
\end{eqnarray}

\noindent Example

We here list $g_{9/2}$ cfp table from Bayman and Lande [3] for the case 
n=4, I=4 and $v$=4.
\bigskip

\begin{center}
\begin{tabular}{|c|c|c|c|c|} \hline
\multicolumn{5}{|c|}{Table 1 (9/2$^{4}$)} \\ \hline
\multicolumn{1}{|c|}{$v_{3}$} & \multicolumn{1}{|c|}{J1} & \multicolumn{1}{|c|}
{$v_{1}$=4} & \multicolumn{1}{|c|}{$v_{2}$=4} & \multicolumn{1}{|c|}{$v_{a}$=4}
 \\ \hline \hline 
\multicolumn{1}{|c|}{3} & \multicolumn{1}{|c|}{3/2} & \multicolumn{1}{|c|}
{-0.080040} & 
\multicolumn{1}{|c|}{0.482164} & \multicolumn{1}{|c|}{-0.12219} \\ \hline 
\hline
\multicolumn{1}{|c|}{3} & \multicolumn{1}{|c|}{5/2} & \multicolumn{1}{|c|}
{0.020346} & 
\multicolumn{1}{|c|}{-0.391860} & \multicolumn{1}{|c|}{-0.05477} \\ \hline 
\hline
\multicolumn{1}{|c|}{3} & \multicolumn{1}{|c|}{7/2} & \multicolumn{1}{|c|}
{-0.620338} & \multicolumn{1}{|c|}{-0.010083} & \multicolumn{1}{|c|}{-0.61704}
 \\ \hline \hline
\multicolumn{1}{|c|}{1} & \multicolumn{1}{|c|}{9/2} & \multicolumn{1}{|c|}
{0.} & \multicolumn{1}{|c|}{0.} & \multicolumn{1}{|c|}{0.} \\ \hline \hline
\multicolumn{1}{|c|}{3} & \multicolumn{1}{|c|}{9/2} & \multicolumn{1}{|c|}
{0.030755} & \multicolumn{1}{|c|}{0.347913} & \multicolumn{1}{|c|}{0.} \\ 
\hline \hline
\multicolumn{1}{|c|}{3} & \multicolumn{1}{|c|}{11/2} & \multicolumn{1}{|c|}
{0.431655} & \multicolumn{1}{|c|}{0.291285} & \multicolumn{1}{|c|}{0.40432} 
\\ \hline \hline
\multicolumn{1}{|c|}{3} & \multicolumn{1}{|c|}{13/2} & \multicolumn{1}{|c|}
{0.594545} & \multicolumn{1}{|c|}{-0.256411} & \multicolumn{1}{|c|}{0.61481} 
\\ \hline \hline
\multicolumn{1}{|c|}{3} & \multicolumn{1}{|c|}{15/2} & \multicolumn{1}{|c|}
{0.205016} & \multicolumn{1}{|c|}{0.505621} & \multicolumn{1}{|c|}{0.15975} 
\\ \hline \hline
\multicolumn{1}{|c|}{3} & \multicolumn{1}{|c|}{17/2} & \multicolumn{1}{|c|}
{0.159916} & \multicolumn{1}
{|c|}{-0.295249} & \multicolumn{1}{|c|}{0.18529} \\ \hline \hline
\end{tabular}
\end{center}
\bigskip
\bigskip

\noindent I = 4 \ \ j = 9/2 \\
\noindent From Table 1 for 4 particles and for the 5 particle listing in 
Bayman-Lande [3] for a J=j, $v$=3 state.
\begin{tabbing}
I = 4 (9/2)$^{4}$ xxxxxxxxxxxx \= xxxxxxxxxxxxx \kill \\
I = 4 (9/2)$^{4}$ \> J = j (9/2)$^{5}$ \\
C($v_{1}$) = 0.030755 \> D($v_{1}$) = -0.018452 \\
C($v_{2}$) = 0.347913 \> D($v_{2}$) = -0.208747 \\
Ratio 11.312 \> Ratio 11.312 \\
\end{tabbing}

It was noted in ref. [4] that in the techniques used to calculate cfp's in 
ref. [3] the two t=4 $v$=4 states ($v_{1}$ and $v_{2}$) were degenerate.  
Hence the emerging $v_{1}$ and $v_{2}$ were somewhat arbitrary.  Any linear 
combination of the two would be equally valid.

\begin{eqnarray}
v_{a} &=& \alpha v_{1} + \beta v_{2} \nonumber \\
v_{b} &=& -\beta v_{1} + \alpha v_{2}
\end{eqnarray}

If instead of a pairing interaction, one uses a seniority conserving 
interaction with non-degenerate levels e.g. the delta interaction, then one 
of the linear combinations, say $v_{a}$ has the interesting property that it 
cannot mix with a $v$=2 state via a seniority non-conserving interaction

\begin{equation}
M = <I=4 v=2 | V | I=4 v_{a}> = 0 
\end{equation}

The cfp's for the state I=4 $v_{a}$ are listed in the last column of Table 1.

Note that both the cfp's for $v_{o}$ = 1 and $v_{o}$ = 3 J=j=9/2 vanish.

Because of the proportionality relationship between 4 particle cfp's and 
5 particle cfp's (2) we have the result 

\begin{equation}
[j^{4} I=4 v_{a} |\} j^{5} J=j v=3] = 0
\end{equation}
At mid-shell the $v$=3 state does not mix with $v$=1 or $v$=5.  This point is 
crucial for the next section.

\section{Further Relations}
We can write

\begin{equation}
M = \sum_{J_{A}} M (J_{A}) E (J_{A})
\end{equation}
where E(J$_{A}$) is the two body matrix element

\begin{equation}
E(J_{A}) = <(j^{2})J_{A} V(j^{2})J_{A}>
\end{equation}

In order to prove that M vanishes for any interaction we must show that

\begin{equation}
M(J_{A}) = 0 \ \ \ J_{A} = 0, 2, 4, 6, 8
\end{equation}

We can get an expression for M in two steps.  First we eliminate one of the 
four particles and get the expression in terms of a three particle 
interaction.  This involves a sum over a three particle angular momentum 
J$_{3}$ and seniority $v_{3}$.  However, only terms with $J_{3} = j$ 
$v_{3}$ = 3 will 
contribute so we supress this index.

We obtain
\begin{eqnarray}
M & = & (4 \times 3/2) \sum_{J_{3}} [j^{3}J_{3} j|\} j^{4} I v=2] \nonumber \\
& & [j^{3} J_{3} j|\} j^{4} I v_{a}=4] <(j^{3})^{J_{3}} V(j^{3})^{J_{3}}>
\end{eqnarray}

The three body matrix element can now be expressed in terms of two body 
matrix elements

\begin{equation}
<(j^{3})J_{3} V(j^{3})J_{3}> = \sum_{J_{A}}[(j^{2})J_{A} j|j^{3} 
J_{3}]^{2} E(J_{A})
\end{equation}

We now note the following n=2 Redmond relation [5] as modified by Zamick 
and Escuderos [6].

\begin{equation}
\sum_{v_{3}} | (j^{2} J_{A}j|\} j^{3} J_{3} v_{3}|^{2} = 1/3 + 2/3 
\{^{j}_{J_{3}} \ ^{j}_{j} \ ^{J_{A}}_{J_{A}}\} (2 J_{A} + 1)
\end{equation}
We finally get

\begin{eqnarray}
M(J_{A}) & = & 6 \sum_{J_{3}} [j^{3}J_{3} j|\}j^{4} I=4 v=2] \nonumber \\
& & [j^{3} J_{3} j|\}j^{4} I=4 v_{a}=4] \nonumber \\
& & [1/3 + 2/3 \{^{j}_{J_{3}} \ ^{j}_{j} \ ^{J_{A}}_{J_{A}}\} (2 J_{A}+1)]
\end{eqnarray}
for $J_{A}$ = 0, 2, 4, 6 and 8.

The ``1/3'' term above will vanish only if the sum $J_{3}$ can be taken over 
all possible values from 3/2 to 17/2.  This is only true for $J_{A}$ = 4 and 
$J_{A}$ = 6.

For $J_{A}$ = 4 we can use a ``4-5'' Redmond relation [5] (also eq. 7 and 
8 of ref. [6]) to obtain

\begin{eqnarray}
M(J_{A}=4) & = & 5 [j^{4} (I=4 v=2),j | \} j^{5} J=j v=3] \nonumber \\
& & [j^{4} (I=4 v_{a}=4),j |\} j^{5} J=j v=3]
\end{eqnarray}
However we have shown in Eq. 5 the following

\begin{equation}
[j^{4} (I=4 v_{a} = 4),j |\} j^{5} J=j v=3] = 0
\end{equation}

Hence we have 
\begin{equation}
M(J_{A}=4)=0
\end{equation}
We can use this fact to prove the result of Eq. (4) i.e. that M = 0.  Recall 
that in the f$_{7/2}$ shell there is no seniority 
mixing for \underline{any} interaction.  One way of proving this is to note 
that one can find three independent seniority conserving interactions.  In the 
f$_{7/2}$ shell there are four identical particles there are four two body 
matrix elements J$_{A}$ = 0, 2, 4 and 6.
However, we can always add an arbitrary constant to get E(J$_{A}$=0)=0, so 
that there are only three.  In the g$_{9/2}$ shell there are four matrix 
elements, J$_{A}$ = 2, 4, 6 and 8.

The three previously mentioned seniority conserving interactions (pairing, 
delta and j(1) $\cdot$ j(2)) will not 
admix seniority $v$=2 and $v$=4 states.  But we have just found a seniority 
violating interaction which will not admix $v$=2 and the $v_{a}$=4 state.  
This latter interaction is 

\begin{equation}
<(j^{2})J_{A} V(j^{2})J_{A}> = V_{o} \delta_{J_{A},4}
\end{equation}

Hence, we can express the four matrix elements in terms of these four 
interactions which will not admix the $v_{a}$=4 state with the $v$=2 state.  
Indeed all M($J_{A}$) vanish.  

Using the same arguments that were given in Eq. (13), (14) and (15) we can 
show that there is no coupling via the interaction (16) (J$_{A}$=4 pairing) 
between the state.  I=4 V$_{a}$=4 and the I=4 v=4 state that is orthogonal 
to it.  This has yet to be shown for a general interaction.

We next consider the energy of the v$_{a}$=4 state.  Using mathematica, 
an expression for the energy was obtained by P. Van Isacker and S. Heinz [7].

\begin{equation}
E ((9/2)^{4}, I=4 v_{a} = 4) \\
= \frac{68}{33} E(2) + E(4) + \frac{13}{15} E(6) + \frac{114}{55} E(8)
\end{equation}
with minor modifications of what has been done up to now, we can explain 
why the coefficient of E(4) is one.  We can write

\begin{equation}
E(j^{4}, I=4 v_{a}=4) = \sum_{J_{a}} G(J_{A}) E(J_{A})
\end{equation}

We can get G(J$_{A}$) from Eq. (12) by replacing the v=2 cfp by the v$_{a}$=4 
cfp.  Hence we obtain 

\begin{equation}
G(J_{A}) = 6 \sum_{J_{A}} [j^{3} J_{3} j|\} j^{4} I = 4 v_{a} = 4]^{2} 
[\frac{1}{3} + \frac{2}{3} \{^{j}_{j_{3}} \ ^{j}_{j} \ ^{J_{A}}_{J_{A}}\} 
(2J_{A}+1)]
\end{equation}

For J$_{A}$=4 we can again use the Redmond relation [5] but care must be taken 
because we now have a diagonal term.  Also the ``1/3'' term now gives a 
contribution.  We obtain

\begin{equation}
G(4) = 6 [1/3 + 2/3 (-1/4)] + vanishing \  cfps = 1
\end{equation}

\section{APPENDIX: The Redmond Recursion Relation in the Seniority Scheme}

We here present the equivalent of the Redmond resursion relation, but for 
cfp's classified by the seniority quantum number $v$ and for which there are 
no redundacies [6]. Here is our formula:

\begin{eqnarray} 
&&(n+1) \sum_{v_s}[j^{n}(v_{0}J_{0})jI_{s}][j^{n}(v_{1}J_{1})jI_{s}|\}
j^{n+1}v_{s}I_{s}] = 
\nonumber\\
&=& \delta_{J_{0}J_{1}}\delta_{v_{0}v_{1}} + n(-1)^{J_{0}+J_{1}} 
\sqrt{(2J_{0}+1)(2J_{1}+1)} \sum_{v_{2}J_{2}} \{^{J_{2}} _{I_{s}} \ ^{j} _{j} \
^{J_{1}} _{J_{0}}\} \nonumber \\
&\times& [j^{n-1}(v_{2}J_{2})jJ_{0}|\}j^{n}v_{0}J_{0}][j^{n-1}(v_{2}J_{2})
jJ_{1}|\}j^{n}v_{1}J_{1}]. 
\end{eqnarray}

Note that we sum over all final seniorities but the final total angular 
momentum is fixed.

I thank Igal Talmi and Piet Van Isacker for helpful discussions.  I thank the 
INT-Seattle where some of the added work was done.

\end{document}